\documentclass[aps,prb,showpacs,twocolumn]{revtex4}
\usepackage{mathrsfs}
\usepackage{amssymb}
\usepackage{amsmath}
\usepackage{bm}
\usepackage{graphicx}
\usepackage{CJK}
\usepackage{natbib}

\newcommand{\sss}[1]{{\scriptscriptstyle #1}}
\begin{document}
\title{Magnetic miniband and magnetotransport property of a
graphene superlattice}
\author{Liwei Jiang}
\author{Yisong Zheng}\email[Author to whom correspondence should be
addressed. Email address: ]{zys@mail.jlu.edu.cn}
\affiliation{National Laboratory of Superhard Materials, Department
of Physics, Jilin University, Changchun 130023, China}

\begin{abstract}
The eigen energy and the conductivity of a graphene sheet subject to
a one-dimensional cosinusoidal potential and in the presence of a
magnetic field are calculated. Such a graphene superlattice presents
three distinct magnetic miniband structures as the magnetic field
increases. They are, respectively, the triply degenerate Landau
level spectrum, the nondegenerate minibands with finite dispersion
and the same Landau level spectrum with the pristine graphene. The
ratio of the magnetic length to the period of the potential function
is the characteristic quantity to determine the electronic structure
of the superlattice. Corresponding to these distinct electronic
structures, the diagonal conductivity presents very strong
anisotropy in the weak and moderate magnetic field cases. But the
predominant magnetotransport orientation changes from the transverse
to the longitudinal direction of the superlattice. More
interestingly, in the weak magnetic field case, the superlattice
exhibits half-integer quantum Hall effect, but with large jump
between the Hall plateaux. Thus it is different from the one of the
pristine graphene.

\end{abstract}
\pacs{68.65.Cd, 71.20.-b, 71.70.Di} \maketitle

\bigskip

\section{introduction}
Since or prior to the experimental acquirement of
graphene\cite{novoselov}, an atomically thin graphitic sheet, its
electronic properties and potential applications were widely
investigated. It was experimentally demonstrated that graphene is a
gapless semiconductor material\cite{novoselov,novoselov2}. In the
Brillouin zone of graphene, there are two inequivalent touching
points between the valence and conduction bands, called the Dirac
points\cite{Zhang}. In the low energy region relative to the Dirac
points, the electron or hole follows a linear dispersion
relation\cite{Zhang,zheng}. Thus, two massless Dirac fermion systems
form in the vicinity of two Dirac points. Such an electron structure
is responsible for most interesting electronic properties unique to
graphene\cite{novoselov2,Zhang,zheng,Katsnelson}. When a
superlattice structure is established on a graphene sheet, the
massless Dirac fermion is subject to a periodic modulation by the
superlattice potential. Such a situation indicates that a graphene
superlattice(GSL) perhaps exhibits unusual electronic
characteristics, unlike both the pristine graphene and the ordinary
semiconductor material\cite{Park1,Park2,Brey,Barbier,Wang}.
Experimentally, many types of GSL have been fabricated and
investigated. For example, electron-beam induced deposition of
adsorbates on graphene membranes creates a dot array with a period
of 5 nm\cite{Meyer}; Epitaxially growth of graphene on some metallic
substrates can form periodic pattern of
supercell\cite{Pletikosi,Diaye,Vazquez,Marchini}. These experimental
progresses certainly make the theoretical studies and device
applications of GSLs realistic.
\par
Accompanying the relevant experimental work, theoretical studies
predicted many interesting electronic properties of GSLs. For
example, some recent theoretical work focused on a simple GSL which
is constructed by exerting a one-dimensional strip-like or
cosinusoidal periodic potential on a graphene
sheet\cite{Park1,Park2,Brey,Park3,Barbier}. It was found that such a
one-dimensional GSL presents multiple Dirac points even in one
valley of the pristine graphene\cite{Park1,Brey}. The dependence of
the number of the Dirac points on the strength and period of the
superlattice potential was analyzed in detail. Furthermore, when a
vertical magnetic field is applied, the low-lying Landau levels(LLs)
become degenerate\cite{Park2}. Accordingly, it was mentioned that
the degeneracy of the LL accounts for the large jump of the Hall
conductivity plateaux of the GSL\cite{Park2,Park3}. Albeit these
existent theoretical results, the electron structure of such a GSL
in the presence of a magnetic field has not yet been comprehensively
revealed. For example, magnetic mini-bands will be established in
the GSL. But such a band structure and its dependence on the
magnetic field strength are yet unknown. The magnetic transport
properties, namely, the diagonal and Hall conductivity spectra of
the GSL, have not been calculated, which may be different from those
of the pristine graphene.
\par
In the present work, by taking the LL states of the pristine
graphene as the basis set, we calculate the energy spectrum of the
GSL in the presence of a magnetic field. We find that the degenerate
LL spectrum is only an electron structure of the GSL in the weak
field limit, namely, in the case that the magnetic length is much
longer than the period of the GSL. On the other hand, when the
magnetic length is smaller than the period of the GSL, the flat LLs
evolve into magnetic minibands with finite dispersion. Furthermore,
if the magnetic length is much smaller than the period of the GSL,
i.e. the strong field case, the superlattice will restore the LL
spectrum of the pristine graphene. As for the magnetotransport
properties, the diagonal conductivity shows notable anisotropy,
especially in the weak magnetic field case. The Hall conductivity
shows indeed the large jumps between the adjacent plateaux in the
low energy region, reflecting the degeneracy of the low-lying LLs.
However, such a quantum Hall effect is destroyed in a stronger
magnetic field when the flat LL spectrum is replaced by the
dispersive magnetic mini-bands.
\par
The rest of the paper is organized as follows: In section II the
theoretical method to solve the electronic eigen-equation of the GSL
in the presence of a vertical magnetic field is briefly elucidated.
Then starting from the Kubo formula the conductivities are
formulated in terms of Green function(GF). In section III, the
numerical results about the spectra of the eigen energy, the density
of states(DOS), the diagonal and Hall conductivities for the GSL are
shown and discussed. Finally, in section IV the main results are
briefly summarized.

\section{Structure and Theory}
When a magnetic field is applied perpendicular to the sheet of a
pristine graphene, the low-energy electron structure in one valley,
say the K valley, can be well described by an effective-mass
Hamiltonian. It takes a form as\cite{zheng}
\begin{equation}
{\cal \hat{H}}_0=\frac{\gamma}{\hbar} \left[\begin{array}{cc}
0 & \hat{\pi}_- \\
\hat{\pi}_+ & 0 \\ \end{array}\right],
\end{equation}
where $\gamma$ is a band parameter, $\hat{\pi}_\pm=\hat{\pi}_x \pm i
\hat{\pi}_y$ with $\hat{\pi}_{\alpha}$ ($\alpha=x, y$) being the
gauged electron momentum. In Landau gauge, one has
$\hat{\pi}_x=\hat{p}_x$ and $\hat{\pi}_y=\hat{p}_y+eBx/c$, with $B$
denoting the magnetic field strength and
$\hat{p}_\alpha=-i\hbar\partial_\alpha$ being the momentum operator.
Noting that K and K' valleys give the same physical result,
hereafter we only treat the electron behavior in K valley. The Dirac
point is taken as the zero energy.
\par
The eigen-equation ${\cal
\hat{H}}_0|nk\rangle=\varepsilon_n|nk\rangle$ can be solved
analytically, which gives the LL spectrum of the pristine
graphene\cite{zheng}. It is
\begin{equation}
\varepsilon_n=\textup{sgn}(n)\hbar \omega_c \sqrt{|n|}, \label{ll}
\end{equation}
and the corresponding wavefunction of the LL state is given by
\begin{equation}
|nk\rangle=\frac{C_n}{\sqrt{L_0}} e^{-iky} \left[\begin{array}{c}
\textup{sgn}(n)(-i){\phi_{|n|-1}} \\ \phi_{|n|}
\end{array}\right].\label{vector}
\end{equation}
In the above two equations, $\hbar \omega_c ={\sqrt{2}\gamma}/l$. It
is defined as the electron cyclotron energy with
$l=\sqrt{c\hbar/eB}$ being the magnetic length. $L_0$ stands for the
linear size of the graphene sheet. And $k$ is electronic wave vector
in the $y$ direction. In addition, the normalization constant is
\begin{equation}
C_n=\left\{\begin{array}{cccc} 1 \quad  && (n=0), \\
1/\sqrt{2} \quad &&(n\neq 0).
\end{array}\right.
\end{equation}
The relevant functions are defined as
\begin{equation}
\textup{sgn}(n)=\left\{\begin{array}{cccc} 1 \quad  && (n>0), \\
0 \quad &&(n=0), \\ -1 \quad  && (n<0),
\end{array}\right.
\end{equation}
and
\begin{equation}
\phi_{|n|}\hspace{-2pt}=\hspace{-3pt}\frac{1}{\sqrt{2^{|n|}|n|!\sqrt{\pi}\hspace{2pt}l}}
\textup{exp}\hspace{-4pt}\left[-\frac{1}{2}\left(\frac{x-x_0}{l}\right)^2\right]\hspace{-3pt}H_{|n|}\hspace{-3pt}\left(\frac{x-x_0}{l}
\right),
\end{equation}
where $x_0=l^2k$, is the center of the electronic harmonic motion;
And $H_n(\xi)$ is the Hermite polynomial.
\par
When we consider the electronic and transport properties of a GSL,
we need to define a one-dimensional periodic potential along $x$
direction, in addition to the Hamiltonian ${\cal \hat{H}}_0$. We
choose the cosinusoidal potential function given by
\begin{equation}
\hat{V}(x)={V_0\over2} \textup{cos}(2\pi x/L).\label{V}
\end{equation}
It can model the periodic gate voltages applied on the graphene
sheet. Thus, the electronic Hamiltonian of such a GSL in the
presence of a vertical magnetic field is given by ${\cal
\hat{H}}={\cal \hat{H}}_0+\hat{V}(x)$. Due to the presence of the
periodic potential, it is impossible to obtain an analytical
solution to the eigen-equation ${\cal \hat{H}}|\lambda
k\rangle=E_\lambda(k)|\lambda k\rangle$. However, we can obtain the
numerical results about it by the exact diagonalization technique.
In doing so, we choose the LL state of the pristine graphene
described above by Eqs(2-6) as the basis set. Then, we can readily
find that the Hamiltonian matrix element is diagonal with respect to
the electron wave vector $k$
\begin{equation}
\langle n'k'|{\cal
\hat{H}}|nk\rangle=\big[\varepsilon_n\delta_{nn'}+\langle
n'k|\hat{V}(x)|nk\rangle\big]\delta_{kk'}.\label{matrix}
\end{equation}
In the above equation, the matrix element of periodic potential can
be worked out analytically. It is given by\cite{Kubo}
\begin{eqnarray}
\langle n'k|\hat{V}(x)|nk\rangle=\frac{V_0C_{n'}C_n}{4}\left[
\textup{sgn}(n'n)\big(J^+_{|n'|-1,|n|-1}+\right.\\\notag
\left.J^-_{|n|-1,|n'|-1}\big)+J^+_{|n'|,|n|}+J^-_{|n|,|n'|}\right],
\end{eqnarray}
where
\begin{eqnarray}
J^\pm_{\sss{NN'}}&=&(\pm i \beta)^{N_1-N_2}
\left[\frac{N_2!}{N_1!}\right]^{1\over 2}\times\\\notag
&&\!\exp\!\!\Big[\pm i{2\pi x_0\over L}-{\beta^2\over
2}\Big]L^{\sss{(N_1-N_2)}}_\sss{{N_2}}(\beta^2),
\end{eqnarray}
with $\beta=\sqrt{2}\pi l/L$, $N_1=\max(N,N')$ and $
N_2=\min(N,N')$. $L_n^{(r)}(\xi)$ stands for the associated Laguerre
polynomials. It is defined as
\begin{equation}
L_n^{(r)}(\xi)=\Big({{e^\xi\xi^{-r}}\over{n!}}\Big){{d^n}\over{d\xi^n}}(e^{-\xi}\xi^{n+r}).
\end{equation}
\par
The eigen wavefunction of the GSL can be formally written as
\begin{equation}
|\lambda
k\rangle=\sum_{n=-\sss{N}_c}^{\sss{N}_c}d_n^\lambda|nk\rangle.
\end{equation}
The expansion coefficients $d_n^\lambda$'s, along with the
corresponding eigen energy $E_\lambda(k)$, can be obtained by
solving the eigen-problem of the Hamiltonian matrix defined above.
In the actual calculation, we need to choose an appropriate cutoff
index $N_c$ which makes the matrix dimension finite. Of course, such
a cutoff should guarantee the stationary results about the low-lying
energy levels.
\par
By defining the retarded GF $G_{\lambda
}(\omega,x_0)=[{\omega-E_\lambda(k)+i\eta}]^{-1}$ with $\eta$ being
a positive infinitesimal, the DOS can be expressed as
\begin{eqnarray}
\rho(\omega)&=&-\frac{1}{\pi
L_0^2}\sum_{\lambda,k}\textup{Im}G_{\lambda }(\omega,x_0)\\\notag
&=&-\frac{1}{2\pi^2l^2L}\sum_\lambda\int_0^L\hspace{-10pt}dx_0
\textup{Im}G_{\lambda}(\omega,x_0).
\end{eqnarray}
According to Kubo's linear response theory, the diagonal and Hall
conductivities of a two-dimensional electron system in the presence
of a magnetic field can be formulated in terms of GF as
\begin{equation}
\sigma_{\alpha\alpha}(\varepsilon)=\frac{e^2\hbar}{\pi
L_0^2}\textup{Tr}\big[\hat{v}_{\alpha}\textup{Im}\hat{G}(\varepsilon)\hat{v}_{\alpha}\textup{Im}\hat{G}(\varepsilon)\big],
\label{Dia}
\end{equation}
and
\begin{eqnarray}
\sigma_{xy}(\varepsilon)=\frac{ie^2\hbar}{\pi L_0^2}\int
_0^\varepsilon\hspace{-8pt}
d\omega\textup{Tr}\big[\hat{v}_{x}\textup{Im}\hat{G}(\omega)
\hat{v}_{y}\frac{d}{d\omega}\textup{Re}\hat{G}(\omega)\\\notag
-\hat{v}_{y}\textup{Im}\hat{G}(\omega)
\hat{v}_{x}\frac{d}{d\omega}\textup{Re}\hat{G}(\omega)\big]
.\label{Ha}
\end{eqnarray}
In the above two formulae, the retarded GF operator is defined as
$\hat{G}(\omega)=[\omega-\hat{\cal{H}}+i\eta]^{-1}$. $\varepsilon$
denotes the chemical potential.
$\hat{v}_{\alpha}=\hspace{-2pt}(i/\hbar)[\hat{H},\hat{\alpha}]=(\gamma/\hbar)\hat{\sigma}_{\alpha}$
is the velocity operator which is associated with the Pauli matrix
$\hat{\sigma}_{\alpha}$. When applying these formulae to the above
GSL, by a straightforward derivation we can obtain the expressions
of the conductivities, given by

\begin{eqnarray}
\sigma_{\alpha\alpha}(\varepsilon)&=&(\frac{e^2}{h})\frac{\hbar^2}{\pi
Ll^2}\int_0^L\hspace{-10pt}dx_0\sum_{\lambda,\lambda'}\\\notag
&&\text{Im}G_{\lambda }(\varepsilon,x_0)
\text{Im}G_{\lambda'}(\varepsilon,x_0)\left|\langle \lambda k|\hat
{v}_\alpha|\lambda' k\rangle\right|^2,\label{diagcdt}
\end{eqnarray}
and
\begin{eqnarray}
\sigma_{xy}(\varepsilon)&=&\hspace{-4pt}(\frac{e^2}{h})\frac{2\hbar^2}{\pi
Ll^2}\int_0^\varepsilon
\hspace{-6pt}d\omega\hspace{-3pt}\int_0^L\hspace{-8pt}dx_0\sum_{\lambda,\lambda'}\text{Im}
G_{\lambda'}(\omega,x_0)\\\notag &&\hspace{-3pt}\text{Re}[G_{\lambda
}(\omega,x_0)]^2 \text{Im}[\langle \lambda k|\hat {v}_x|\lambda'
k\rangle\langle \lambda k|\hat {v}_y|\lambda' k\rangle].
\end{eqnarray}
The matrix elements of the velocity operators involved in the above
two expressions are given by
\begin{eqnarray}
\langle \lambda k|\hat {v}_x|\lambda'
k\rangle&=&\hspace{-3pt}-i{\gamma\over\hbar}\sum_{m,n}d_m^{\lambda'}d_n^{\lambda}C_mC_n\\\notag
&&\hspace{-3pt}\left[\text{sgn}(m)\delta_{|m|-1,|n|}
-\text{sgn}(n)\delta_{|m|,|n|-1}\right]\label{element1},
\end{eqnarray}
and
\begin{eqnarray}
\langle \lambda k|\hat {v}_y|\lambda'
k\rangle&=&\hspace{-3pt}{\gamma\over\hbar}\sum_{m,n}d_m^{\lambda'}d_n^{\lambda}C_mC_n\\\notag
&&\hspace{-3pt}\left[\text{sgn}(m)\delta_{|m|-1,|n|}
+\text{sgn}(n)\delta_{|m|,|n|-1}\right]\label{element2}.
\end{eqnarray}
We can readily prove that the two relations, $\langle \lambda
k|\hat{v}_x|\lambda k\rangle=0$ and $\langle \lambda
k|\hat{v}_y|\lambda k\rangle=(1/\hbar)\partial E(k)/\partial k$,
always hold true for any state. These relations are useful for us to
analyze the numerical results about the magnetotransport property.
\section{Numerical results}
We are now in a position to perform the numerical calculations about
the electronic and transport properties of the GSL in the presence
of a magnetic field, according to the theoretical approach given
above. First of all, we calculate the electronic eigen energy
spectrum for the cases of different magnetic fields. We take the
cutoff index $N_c=1200$ which ensures the reliable result about the
low-lying eigen energies close to the Dirac point. These spectra of
a GSL are shown in Fig.1. The relevant parameters of the GSL satisfy
the relation $V_0=6\pi \varepsilon_{\hspace{-2pt}_L}$ with
$\varepsilon_{\hspace{-2pt}_L}=\gamma/L$. According to previous
work\cite{Park2,Park3}, such a GSL displays triple Dirac points in
one valley of the pristine graphene in the case of zero magnetic
field. We reproduce such a result by choosing the plane waves as the
basis set to establish the Hamiltonian matrix of the GSL in the
absence of a magnetic field. Such a dispersion relation is shown in
Fig.1(a). In Fig.1(b) some low-lying eigen energies are shown as
functions of the electron wavevector $k$(rescaled as $x_0$) for a
relatively weak magnetic field. The corresponding magnetic length is
larger than the period of the GSL. We can see that these eigen
energies scarcely depend on $x_0$. This is a reasonable result,
because that the matrix elements of the Hamiltonian hardly depend on
the central position of the LL states when the magnetic length is
much larger than the period of the GSL. Hence the eigen energies
form the flat magnetic minibands. These flat minibands shown in
Fig.1(b) can be viewed as the LL spectrum of the GSL. However, these
levels are not exactly dispersionless, which can be clearly seen
from the insets which blow up the weak dispersion. Besides, the
level distribution in Fig.1(b) is different from the LL spectrum of
the pristine graphene. We can see from this figure that the three
energy levels at the Dirac point are nearly degenerate. So are the
two groups of the levels near the Dirac points. These results verify
the LL degeneracy in the GSL concluded in previous work by using a
distinct theoretical approach\cite{Park2,Park3}. When the magnetic
field increases, the flat LL spectrum changes into dispersive
magnetic mini-bands. Such a situation is shown in Fig.1(c) where the
magnetic length is smaller than the period of the GSL by one order
of magnitude. Hereafter we call such a case, namely, the occurrence
of the dispersive magnetic mini-bands, as the moderate field case.
Fig.4(d) shows the eigen energy spectrum when the magnetic length is
far smaller than the period of the GSL. We can see that such a
strong magnetic field restores the eigen energy spectrum to the LL
spectrum of the pristine graphene, described by Eq.(\ref{ll}). This
is because that in such a strong field case the periodic potential
of GSL can be viewed as a small perturbation which only gives a
trivial correction to the LL spectrum of the pristine graphene.
\begin{figure}
\begin{center}
\scalebox{0.36}{\includegraphics{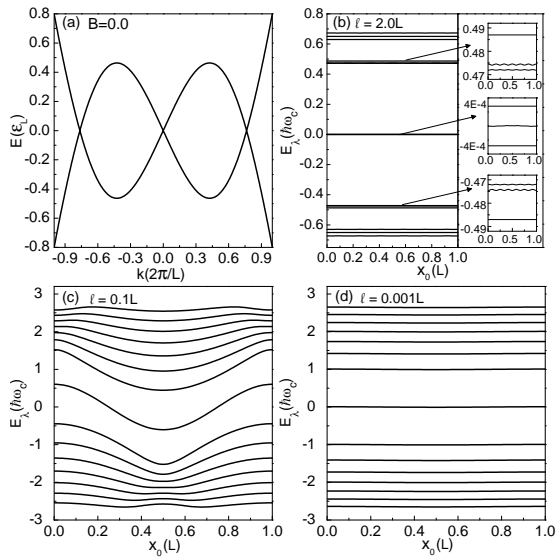}} \caption{Low-lying eigen
energies versus electronic wavevector k (rescaled as $x_0$ when a
magnetic field is present) of a GSL with $V_0=6\pi
\varepsilon_\sss{L}$ and $\varepsilon_\sss{L}=\gamma/L$. (a) The
dispersion relation in the zero magnetic field case. The electronic
wavevector along $x$ direction is fixed at $k_x=0$. (b) The weak
field case, i.e. $l=2.0L$. The insets show the weak dispersion of
the nearly degenerate levels in detail.(c) The moderate field case
with $l=0.1L$ and (d) the strong field case with $l=0.001L$.}
\end{center}\end{figure}
\par
In order to see the influence of the magnetic field on the eigen
energy spectrum of the GSL more clearly, in Fig.2 we show the eigen
energies as functions of the magnetic length for a specific
wavevector at $x_0=0$. We can see that in the moderate field regime
the eigen energies depend sensitively on the variation of the
magnetic field. In addition, in the weak field limit all the LLs
tend to triply degenerate, which follows the triple Dirac points of
the GSL. In Fig.\ref{dos} we show the calculated DOS for such a GSL.
The weak field case is shown in Fig.\ref{dos}(a). The sharp peaks in
such a DOS spectrum correspond to LLs as shown in Fig.1(b). Noting
that the three high peaks near the Dirac point arise from the triple
degeneracy of these LLs. In the moderate field case, as shown in
Fig.\ref{dos}(b), the DOS spectrum exhibits a series of peaks. These
peaks indicate the singularities in the DOS, which correspond to the
minima and maxima of the magnetic mini-bands as shown in Fig.1(c).
In the strong field case shown in Fig.3(c), the DOS spectrum is
almost the same as that of the pristine graphene in the presence of
a magnetic field.
\begin{figure}
\begin{center}
\scalebox{0.34}{\includegraphics{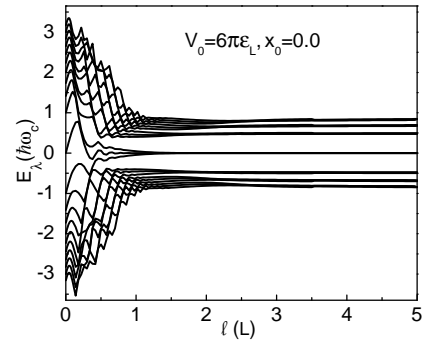}} \caption{Low-lying eigen
energies versus the magnetic length of a GSL with $V_0=6\pi
\varepsilon_\sss{L}$ at $x_0=0$.}
\end{center}\end{figure}
\begin{figure}
\begin{center}
\scalebox{0.32}{\includegraphics{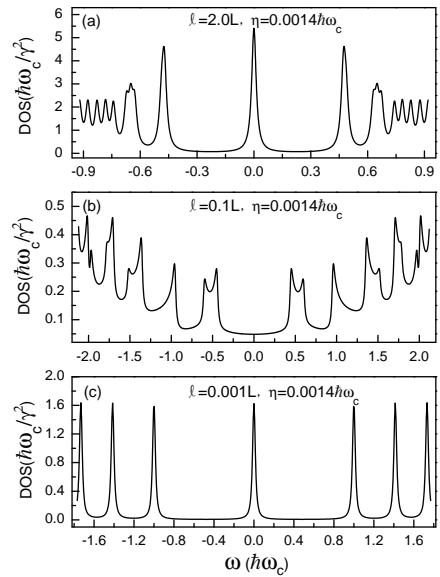}} \caption{The DOS
spectrum of the GSL for (a) the weak, (b) the moderate and (c) the
strong magnetic field cases, respectively. \label{dos}}
\end{center}\end{figure}
\par
\begin{figure}
\begin{center}
\scalebox{0.4}{\includegraphics{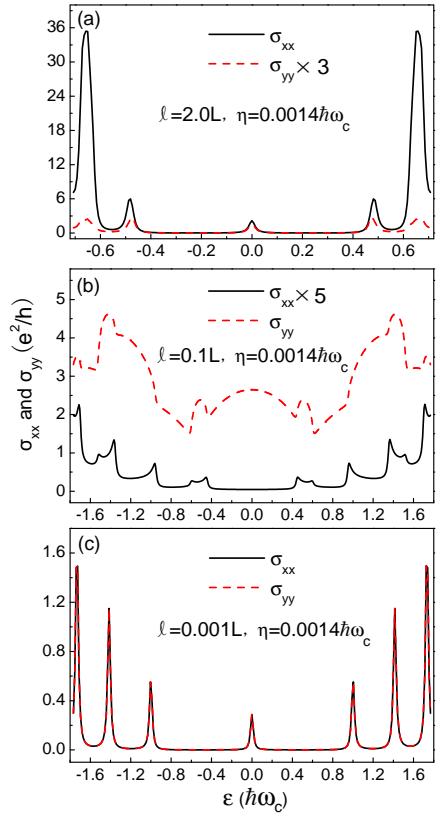}} \caption{(Color online)
The diagonal conductivities $\sigma_{xx}$ and $\sigma_{yy}$ versus
the chemical potential for a GSL with $V_0=6\pi \varepsilon_\sss{L}$
for some typical magnetic fields. (a) Magnetic length $l=2.0L$, (b)
$l=0.1L$, and (c) $l=0.001L$. Noting that $\sigma_{yy}$ in (a) is
multiplied by 3 and $\sigma_{xx}$ in (b) is multiplied by 5, in
order to show these curves clearly.\label{cdtxx}}
\end{center}\end{figure}

Now we turn to study the magnetotransport properties of the GSL. The
diagonal conductivity is calculated as a function of the chemical
potential. And the calculated results are shown in Fig.\ref{cdtxx}.
From Fig.\ref{cdtxx}(a) we can see that in the weak field case
$\sigma_{xx}$ is larger than $\sigma_{yy}$. It indicates that the
diagonal conductivity shows a strong anisotropy. We have known that
the average velocity $\langle \lambda k|\hat{v}_x|\lambda
k\rangle=0$ for any eigen-state. Besides, in the weak field case the
average velocity along $y$ direction is ignorably small, namely,
$\langle \lambda k|\hat{v}_y|\lambda k\rangle=(1/\hbar)\partial
E_\lambda(k)/\partial k\approx 0$, due to the flat band structure as
shown in Fig.1(b). Therefore, the diagonal conductivity in the weak
field case is determined by the interband contributions. From the
numerical calculation we find that the matrix elements $\langle
\lambda k|\hat{v}_x|\lambda' k\rangle$ are notably larger than
$\langle \lambda k|\hat{v}_y|\lambda' k\rangle$ for any two distinct
eigen-states. As a result, the conductivity in $x$ direction is
larger than that in $y$ direction in the weak field case. In fact,
the anisotropy of the diagonal conductivity arises from the
anisotropic band structure around each Dirac points of the GSL.
Previous work reported that the band structure of the GSL is notably
anisotropic. Therefore, $\langle \hat{v}_x \rangle
>\langle \hat{v}_y \rangle$ and $\sigma_{xx}>\sigma_{yy}$ holds
true in zero magnetic field case\cite{Barbier}. We can infer that
such an anisotropic band structure at each Dirac point of the GSL
also controls the anisotropy in the electronic magnetotransport
process presented here. The anisotropy in the diagonal conductivity
remains in the moderate field case. However, opposite to the weak
field case, the diagonal conductivity $\sigma_{xx}$ becomes the
smaller one. In Fig.\ref{cdtxx}(b) we can see that $\sigma_{yy}$ is
nearly 10 times $\sigma_{xx}$. Moreover, the peaks of $\sigma_{xx}$
just correspond to the dips of $\sigma_{yy}$. We can explain these
features in the following way. The magnetotransport properties of
the GSL in the moderate field case are determined by the dispersive
magnetic mini-bands. Then the average velocity along $y$ direction
takes a nonzero value in any mini-band, except at the band-edges.
Namely, $\langle \hat{v}_y\rangle=\langle \lambda
k|\hat{v}_y|\lambda k\rangle=(1/\hbar)\partial E_\lambda(k)/\partial
k\neq 0$. In contrast to it, the average value of the velocity along
$x$ direction is always equal to zero, i.e. $\langle \lambda
k|\hat{v}_x|\lambda k\rangle=0$ in any state. Such a result means
that $\sigma_{yy}$ has an intraband contribution, but the
$\sigma_{xx}$ does not. From Eq.(\ref{diagcdt}), the formula about
the diagonal conductivity, we can see that $\sigma_{yy}$ is mainly
determined by the intraband term which is proportional to $|\langle
\lambda k|\hat{v}_y|\lambda k\rangle|^2$. At the minimum and the
maximum of a specific mini-band, one has $\langle \lambda
k|\hat{v}_y|\lambda k\rangle=0$. This indicates that the dips in the
$\sigma_{yy}$ spectrum occur at these positions. On the other hand,
the conductivity $\sigma_{xx}$ lacks the intraband contribution.
Furthermore, the interband terms depend on the DOS of the magnetic
mini-bands involved. This means that the conductivity peaks in the
$\sigma_{xx}$ spectrum occur at the singularities of the DOS. And
these singularities just correspond to the minimum or the maximum of
a mini-band. Fig.\ref{cdtxx}(c) shows the diagonal conductivity
spectra in the strong field case. We can see that the anisotropy
disappears. And the conductivity spectrum is very similar to that of
pristine graphene.
\par
\begin{figure}
\begin{center}
\scalebox{0.4}{\includegraphics{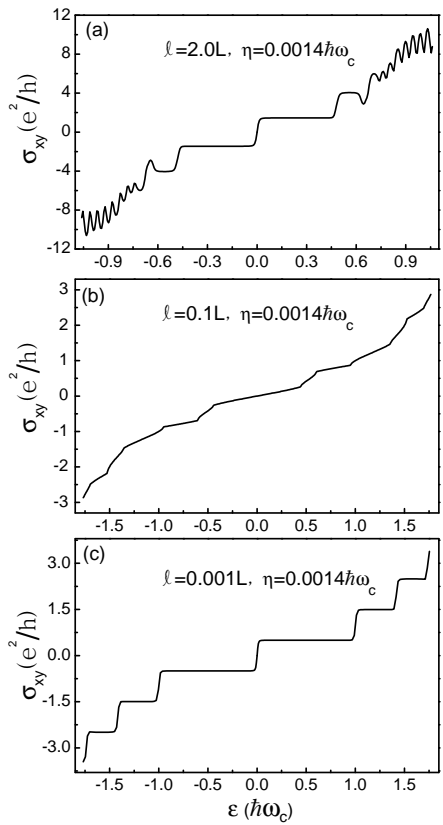}} \caption{The Hall
conductivity $\sigma_{xy}$ versus the chemical potential for a GSL
with $V_0=6\pi \varepsilon_\sss{L}$ for some typical magnetic
fields. (a) Magnetic length $l=2.0L$, (b) $l=0.1L$, and (c)
$l=0.001L$. \label{hallc}}
\end{center}\end{figure}

The spectra of the Hall conductivity are shown in Fig.\ref{hallc}
for the typical magnetic fields. At first, from Fig.\ref{hallc}(a)
we can see that in the weak field case the GSL exhibits the quantum
Hall effect. But the resolvable Hall plateaux only appear at the
positions: $\pm3/2(e^2/h), \pm9/2(e^2/h)$ near the Dirac point. This
situation is different from the half-integer quantum Hall effect of
the pristine graphene. In fact, such a quantum Hall effect just
follows the degenerate LL spectrum (strictly speaking, the nearly
degenerate LLs) in the weak field case as shown in Fig.1(b). The
triply degenerate low-lying LLs and the conduction-valence band
symmetry of the energy spectrum can well account for these plateaux
shown in Fig.\ref{hallc}(a). When the chemical potential goes away
from the Dirac point, the dense LLs bring about the oscillating
character in the Hall conductivity spectrum, instead of the plateau
structure in the low energy region. In the moderate field case the
flat LL spectrum disappears. As a result, Hall conductivity does not
exhibit the well-defined quantum Hall plateaux. Such a result is
shown in Fig.\ref{hallc}(b). However, when the magnetic field
increases to the strong field regime, the GSL exhibits the
half-integer quantum Hall effect of the pristine graphene which is
shown in Fig.\ref{hallc}(c).

\section{summary}
We have studied the magnetic mini-band structure and the
magnetotransport properties of a GSL formed by applying a
one-dimensional cosinusoidal potential on a graphene sheet. At
first, by using the LL states of the pristine graphene as the basis
set and by solving the Hamiltonian matrix of the GSL, we have
obtained the eigen energy spectrum of the GSL subject to a magnetic
field. We found that the GSL exhibits three different electronic
structures as the magnetic field increases. When the magnetic length
is larger than the period of the GSL, namely, the weak field case,
the GSL presents flat magnetic mini-bands near the Dirac point.
These flat mini-bands can be viewed as the LL spectrum of the GSL.
Unlike the LL spectrum of the pristine graphene, these LLs of GSL
are nearly degenerate with three LLs as a group. And in the weak
field limit the LL spectrum tends to triply degenerate. When the
magnetic length is smaller than the period of the GSL by one order
of magnitude, namely, the moderate field case, the flat LL spectrum
in the weak field case is replaced by the dispersive mini-band
structure. And when the magnetic field increases further to enter
the strong field regime, namely, the magnetic length is far shorter
than period of the GSL, the eigen energy spectrum gets back to the
LL spectrum of the pristine graphene. Corresponding to these
different electronic structures, the diagonal and Hall
conductivities present different characteristics in different
magnetic field regions. In the weak and moderate field cases, the
diagonal conductivity exhibits a strong anisotropy. However, the
predominant magnetotransport orientation changes from the transverse
to the longitudinal direction of the superlattice when the magnetic
field increases from the weak to the moderate case. More
interestingly, in the weak field case the GSL presents large jumps
between the Hall plateaux, different from the quantum Hall effect of
the pristine graphene. This feature arises from the degenerate LL
spectrum of the GSL in the weak field case.

\section{Acknowledgements\label{Acknowledgements}}
 This work was financially supported by the National Nature Science
Foundation of China under Grant No. NNSFC10774055, the Specialized
Research Fund for the Doctoral Program of Higher Education (Grant
No. SRFDP20070183130), and National Found for Fostering Talents of
Basic Science (Grant No. J0730311).

\clearpage

\end{document}